\documentclass[11pt,a4paper]{article}

\usepackage[utf8]{inputenc}
\usepackage[T1]{fontenc}
\usepackage{newpxtext,newpxmath} 
\usepackage{microtype}         
\usepackage{url}
\usepackage{algorithm}
\usepackage{algpseudocode} 
\usepackage{booktabs}
\usepackage{subcaption}
\usepackage{caption}
\usepackage{float}
\usepackage{amsmath}
\usepackage{bm}
\usepackage{makecell}
\usepackage{appendix}
\usepackage{titlesec}

\usepackage{geometry}
\usepackage{parskip}           

\usepackage{xcolor}
\definecolor{abstractbg}{RGB}{235,243,250} 
\definecolor{titlecolor}{RGB}{20,20,40}      

\usepackage{tcolorbox}
\tcbset{
  fullwidthabstract/.style={
    width=\textwidth,
    colback=abstractbg,
    colframe=abstractbg,
    boxrule=0pt,
    arc=4pt,
    left=10pt,
    right=10pt,
    top=10pt,
    bottom=10pt,
  },
}

\newcommand{\companylogo}{\includegraphics[height=1.5cm]{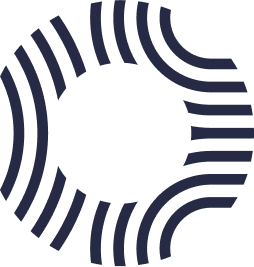}}

\usepackage[
natbib=true,
backend=biber,
style=numeric,
sorting=ynt
]{biblatex}

\begin{document}

\vspace*{-1cm}
\begin{tcolorbox}[fullwidthabstract, title={\vspace{3mm} \companylogo\quad \huge \color{titlecolor}{\hspace{0.1cm} Orb-v3: atomistic simulation at scale}}]

\begin{center}
Benjamin Rhodes\footnotemark[1], Sander Vandenhaute\footnotemark[1], Vaidotas Šimkus, James Gin,\\ Jonathan Godwin, Tim Duignan, Mark Neumann\\[1em] 
\texttt{\{ben,mark\}@orbitalmaterials.com} \\
Orbital Materials \\[1.5em]

\end{center}

We introduce Orb-v3, the next generation of the Orb family of universal interatomic potentials. Models in this family expand the performance-speed-memory Pareto frontier, offering near SoTA performance across a range of evaluations with a $\geq10\times$ reduction in latency and $\geq8\times$ reduction in memory. Our experiments systematically traverse this frontier, charting the trade-off induced by \emph{roto-equivariance}, \emph{conservatism} and \emph{graph sparsity}. Contrary to recent literature, we find that non-equivariant, non-conservative architectures can accurately model physical properties, including those which require higher-order derivatives of the potential energy surface. \vspace{1em}

This model release is guided by the principle that the most valuable foundation models for atomic simulation will excel on all fronts: accuracy, latency and system size scalability. The reward for doing so is a new era of computational chemistry driven by high-throughput and mesoscale all-atom simulations.

\footnotetext{*equal contribution}
\end{tcolorbox}

\vspace{1.5em}

Simulation-based computational chemistry is undergoing a remarkable transition. For several decades, the field has relied on the success of density functional theory (DFT) \cite{kohn_sham} and other approximate solutions to the Schrödinger equation—a framework that has unlocked unprecedented insights into the electronic structure and physical properties of matter. However, the computational cost, typically scaling as $O(N^3)$ or more, is prohibitive for large systems and has become a bottleneck that limits the use of DFT in high-throughput predictive simulations. Universal Machine Learning Interatomic Potentials (MLIPs) represent a new paradigm, promising \textit{ab initio} accuracy for a wide range of chemistries at enlarged spatio-temporal scales.

MLIP design is broadly composed of two tracks. The first track is concerned with \emph{universality}; how can we learn an accurate single potential for all chemical systems? This requires large-scale dataset creation efforts \cite{jain2013commentary, alexandria, barrosoluque2024openmaterials2024omat24, kaplan2025matpes, mazitov2025petmad}, model-building \cite{deng2023chgnetpretraineduniversalneural, batatia2023macehigherorderequivariant, yang2024mattersim, park2024sevennet, neumann2024orb, barrosoluque2024openmaterials2024omat24, bochkarev2024grace, fu2025learning, amin2025towards} and rigorous evaluations \cite{riebesell2024matbenchdiscoveryframework, loew2024mdr, wines2024chips, MLIPX,  kaplan2025matpes, mazitov2025petmad}. The second track is concerned with \emph{scalability}; how can we model realistic systems in some of the most important applications -  bio-materials, chemical reactions or enzymatic processes? This requires more efficient all-atom architectures \cite{neumann2024orb, pelaez2024torchmd, kaplan2025matpes} and coarse-grained potentials \cite{Majewski2023, Wellawatte2023}. A grand challenge for the field is to unite these two tracks, and deliver a universal model, usable by material scientists and biochemists alike, that can accurately simulate novel systems across several orders of spatio-temporal magnitude.

\begin{figure}[!ht]
    \centering
    \includegraphics[width=0.975\linewidth]{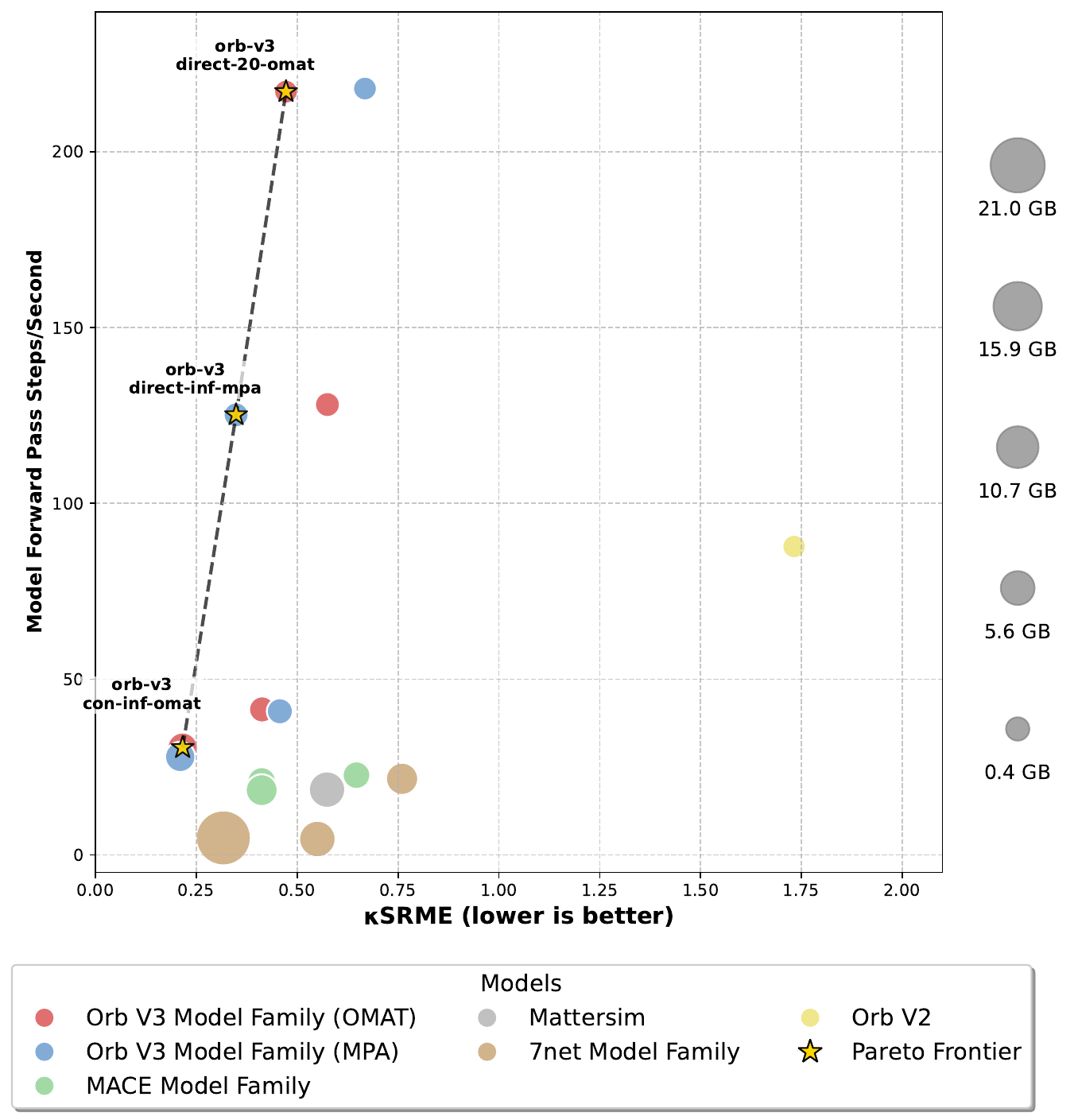}
    \caption{The Pareto frontier for a range of universal Machine Learning Interatomic Potentials. The $K_{SRME}$ metric assesses a model's ability to predict thermal conductivity via the Wigner formulation of heat transport \cite{pota2024thermal} and requires accurate geometry optimizations as well as second and third order derivatives of the PES (computed via finite differences). The y-axis measure a model's forward passes per second on a dense periodic system of 1000 atoms, disregarding graph construction time, measured on a NVIDIA H200. Point sizes represent max GPU memory usage. Y-axis jitter (+/- 5 steps/second) has been applied to allow visualization of overlapping points. Model families include a range of specific models with broadly the same architecture, but may be different sizes or trained on different datasets. More details are provided in Appendix \ref{sec:model_families}.}
    \label{fig:pareto}
\end{figure}

In this technical report, we introduce the Orb-v3 series of models: \emph{universal} and \emph{scalable} all-atom models at various points on the performance-speed-memory Pareto frontier. At one end of this spectrum are smooth, conservative potentials with a high degree of roto-equivariance induced by a new gradient-based regularization scheme called equigrad. Such models excel in performance, predicting vibrational, thermodynamic and mechanical properties with high precision. At the other end of the spectrum are non-conservative models with a sparser atomic graph featurization. As shown in Figure \ref{fig:pareto}, such models are highly scalable---often more than 10x faster and with 8x lower memory footprint than alternative MLIPs, whilst still enjoying excellent performance when trained on large ab initio molecular dynamics (AIMD) datasets such as OMAT24.

\newpage
\section*{Orb-v3 Models}
\label{sec:model_naming}
Orb-v3 is a family of models that share the same basic architecture as Orb-v2 \cite{neumann2024orb, SanchezGonzalez2020LearningTS} as well as the same  diffusion pretraining scheme. Despite this similar top-level training strategy, we find that there is a range of often subtle design choices that affect a model's performance. We enumerate the full list of these in Appendix \ref{sec: orbv3 modelling updates}, focusing here on the three most significant choices: \emph{conservatism}, \emph{maximum neighbor limits} and \emph{choice of dataset}.

These three key variables chart a path across the performance-speed-memory Pareto frontier. Thus, our publicly released models\footnote{Available under an Apache 2.0 License at \url{https://github.com/orbital-materials/orb-models}.} use suffixed names of the form \texttt{orb-v3-X-Y-Z}, where
\begin{align*}
 \text{X} \in \{\texttt{direct}, \texttt{conservative}\}, \quad \text{Y} \in \{\texttt{20}, \texttt{inf}\}, \quad \text{Z} \in \{\texttt{omat}, \texttt{mpa}\}
\end{align*}
where X denotes whether forces and stress are computed as gradients of the energy, Y refers to a maximum number of neighbors per atom, and Z is the final dataset that a model was trained on.

\subsection*{Conservatism and equigrad}
Orb-v2 demonstrated that non-conservative potentials can be fast, low-memory and performant. However, as shown by \citet{bigi2024dark}, they may have inherent limitations such as not conserving energy in NVE molecular dynamics. We argue that the choice of direct versus conservative models may ultimately be workflow-dependent, and thus release both types.

During training, our conservative models benefit from \emph{equigrad}, a new roto-equivariance-inducing regularization scheme.
The key insight is that we can quantify and improve the rotational invariance of the energy prediction by regularizing the gradient of $E$ with respect to an identity rotation matrix applied to atomic positions. See the corresponding Section below for more information.

\subsection*{Neighbor limits}
Orb-v2 defined atomic neighborhoods by a max radius of 10 \r{A} and a limit of 20 neighbors. We have since discovered that neighbor limits come with performance penalties for certain calculations---likely due to the discontinuities they create in the PES---corroborating the findings of \citet{fu2025learning}. \emph{However}, unlike \citet{fu2025learning}, we still release models that use neighbor limits, because they occupy a different part of the performance-speed-memory Pareto frontier. 

\subsection*{Datasets and distillation}
The OMat24 dataset \citep{barrosoluque2024openmaterials2024omat24} has quickly become the default dataset for universal MLIPs \citep{park2024sevennet}. Roughly half of its 100M datapoints come from AIMD, and the other half from `rattling' existing low-energy structures. Early in development, we found that these rattled systems had deleterious effects on our models when evaluated on out-of-distribution hetero-diatomic systems (see Appendix \ref{sec: filtering omat}). Thus, all \texttt{orb-v3-*-omat} models are only trained on the AIMD subset of OMat24.

We also release models trained on \texttt{mpa}, which is shorthand for the combination of MPTraj \cite{jain2013commentary} and Alexandria (PBE) \citep{alexandria}. These datasets have been instrumental in the development of universal MLIPs, but in our view have now been supplanted by OMat24, which is much larger, more diverse in terms of off-equilibrium structures, and uses newer pseudo-potentials. We release \texttt{mpa} models for compatibility with existing benchmarks such as Matbench-Discovery, but advise users of \texttt{orb-v3} to default to the \texttt{-omat} versions.

During development, we observed that our \emph{direct} Orb-v3 models---which have more degrees of freedom and are thus more data-dependent---tend to overfit to forces when trained on \texttt{mpa}, and struggle to accurately model second- and third-order derivatives of the PES. This problem occurs even when finetuning on \texttt{mpa} from an \texttt{-omat} base model. Intriguingly, we were able to resolve this problem via a simple form of \emph{distillation} of conservative models into direct models. Concretely, we used \texttt{orb-v3-conservative-inf-mpa} to generate a static set of energy, force and stress predictions across the entirety of \texttt{mpa}, and then used those predictions as targets when training \texttt{orb-v3-direct-*-mpa} models. See Appendix \ref{sec: distillation direct mpa} for further discussion.

\begin{figure}[!htbp]
    \vspace{-1em}
    \centering
    \includegraphics[width=\linewidth]{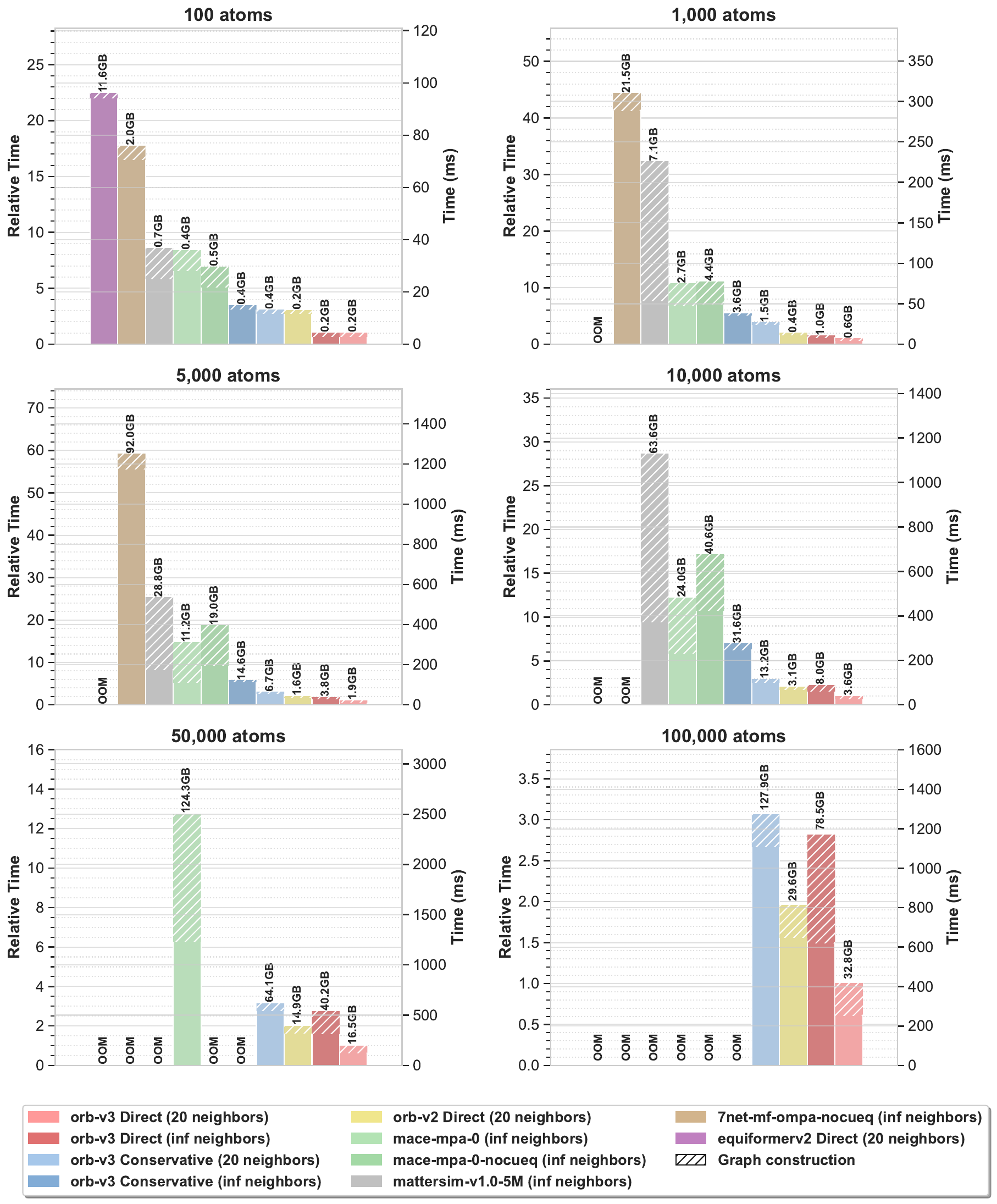}
    \caption{\small{Speed + max GPU memory allocated on an NVIDIA H200 for the computation of energies, forces and stress. The batch size is fixed to 1, but we vary the number of atoms across the subplots. Relative times are computed with respect to the fastest model: \texttt{orb-v3 Direct (20 neighbors)}. Times include both model inference and graph construction, with the latter marked by hatched lines. The graph construction method for Orb is a function of the number of atoms, as described in Appendix \ref{sec: efficient graph construction}. A key takeaway from this figure is that extreme scalability requires a confluence of i) efficient graph construction ii) Finite max neighbors iii) Non-conservative direct predictions. For the baselines, we use \texttt{mace-medium-mpa-0} (\texttt{v0.3.10}, \texttt{cuequivariance-torch v0.1.0}), \texttt{mattersim-v1.0.0-5m} (\texttt{v1.1.2}), \texttt{7net-mf-ompa} (\texttt{v0.11.0}). All models are benchmarked using \texttt{PyTorch v2.6.0+cu124}. Alternative libraries, like JAX, may yield further improvements for some models, but is out of scope for this work.}}
    \label{fig:profiles}
\end{figure}

\section*{Speed and Memory}
\label{section: speed_and_memory}
Molecular dynamics simulations are typically run using time steps on the order of a femtosecond, and yet many physically interesting phenomena only emerge at the nanosecond scale or beyond. This entails making millions of sequential calls to an MLIP to iteratively update atomic positions.

As shown in the Pareto frontier plot of Figure \ref{fig:pareto}, \texttt{orb-v3-direct-*} are the \emph{only} universal MLIPs that can compute hundreds, rather than tens, of forward passes per second, thereby passing the threshold of one million steps per hour for small systems. This step-change in speed, at a relatively low cost in accuracy, makes \texttt{orb-v3-direct-*} models powerful tools for accelerated scientific discovery.

Another clear trend from Figure \ref{fig:pareto} is the memory efficiency of \texttt{orb-v3-direct-*} models. In order to stress test memory efficiency (and latency), Figure \ref{fig:profiles} profiles a range of MLIPs on periodic systems of up to 100,000 atoms. All baseline methods, as well as our conservative models, encounter Out Of Memory (OOM) errors for 100,000 atoms; in contrast, \texttt{orb-v3-direct-20} uses only 32.8GB of GPU memory and completes in under half a second.

Finally, it is interesting to observe in Figure \ref{fig:profiles} that state-of-the-art MLIPs are easily bottlenecked by expensive graph construction routines which can dominate their runtime. As explained in Appendix \ref{sec: efficient graph construction}, we have prioritized efficient off-the-shelf solutions using a combination of brute force and GPU-accelerated nearest neighbors routines, via the \texttt{cuML} library \citep{raschka2020machine}.

\section*{Benchmark Results}
\label{section: results}

In order to evaluate the performance of the models along the Pareto frontier defined by the Orb-v3 family of models, we use several well established benchmarks which incorporate tasks covering a wide variety of computational workflows, including geometry optimization, phonon calculations, and molecular dynamics.

\subsection*{Matbench Discovery}
Table \ref{matbench_table} reports F1 and $\kappa_{\text{SRME}}$ from the Matbench-Discovery benchmark \citep{riebesell2024matbenchdiscoveryframework}. F1 is a metric that assesses a model's thermodynamic stability predictions and requires accurate geometry optimizations combined with single-point energy calculations (relative to a pre-existing convex hull). The $\kappa_{\text{SRME}}$ metric assesses a model's ability to predict thermal conductivity via the Wigner formulation of heat transport \cite{pota2024thermal} and requires accurate geometry optimizations as well as second- and third-order energy derivative estimation via finite differences. In addition, we report model forward passes per second, giving a sense of the tradeoffs available at various levels of benchmark performance. Particularly of note is the performance of Orb-v3 models when used for computing thermal conductivity, demonstrating that it is possible to train rotationally non-invariant, direct models which yield competitive results (and by implication, admit smooth second- and third-order derivatives of the potential energy surface).

\begin{table}[!t]
    \centering
\setlength{\tabcolsep}{9.5pt} 
\begin{tabular}{lccc}
\toprule
Model & F1 ↑ & $\kappa$SRME ↓  & \thead{Steps/Second\\(1k atoms) ↑} \\
\midrule
eSEN-30M-OAM \cite{fu2025learning} & 0.925 & 0.170 & --- \\
SevenNet-MF-ompa \cite{park2024sevennet} & 0.901 & 0.317 & 3.5 \\
GRACE-2L-OAM \cite{bochkarev2024grace} & 0.880 & 0.294 & --- \\
MACE-MPA-0 \cite{batatia2024foundationmodelatomisticmaterials} & 0.852 & 0.412 & 21.2 \\
DPA3-v2-OpenLAM \cite{Zeng2025DeePMDkitV3} & 0.890 & 0.687 & --- \\
MatterSim v1 5M \cite{yang2024mattersim} & 0.862 & 0.574 & 18.8 \\
eqV2 M \cite{liao2023equiformerv2} & 0.917 & 1.771 & OOM \\
ORB v2 \cite{neumann2024orb} & 0.880 & 1.732 & 88.3\\
\midrule
Orb-v3-direct-20-mpa & 0.877 & 0.668 & 216.5\\
Orb-v3-direct-inf-mpa & 0.883 & 0.348 & 125.0 \\
Orb-v3-conservative-20-mpa & 0.902 & 0.457 & 41.2\\
Orb-v3-conservative-inf-mpa & 0.906 & 0.210 & 28.1\\
\midrule
Orb-v3-direct-20-omat & --- & 0.472 & 216.5\\
Orb-v3-direct-inf-omat & --- & 0.575 & 125.0\\
Orb-v3-conservative-20-omat & --- & 0.413 & 41.2\\
Orb-v3-conservative-inf-omat & --- & 0.216 & 28.1\\
\bottomrule
\end{tabular}
\caption{Matbench results for a range of Orb-v3 models. Orb-v3 models perform competitively, whilst having significantly improved speed and memory profiles. Note that results for \texttt{*-omat} models on the discovery portion of the benchmark are not included, as OMat24 uses PBE54 VASP pseudopotentials, making them incompatible with the WBM test set. See Appendix \ref{sec:compatability} for an analysis of how these datasets result in broadly similar potentials.}

\label{matbench_table}

\end{table}

\subsection*{Physical Property Predictions}
\label{section: phonons}
Ultimately, the goal of developing general purpose MLIPs is to enable efficient and high-fidelity predictions of materials properties at scale. Benchmark performance on relative targets, such as F1 with respect to a predefined energy hull, does not necessarily transfer into accurate and reliable prediction of physical properties; this is well demonstrated by the new Matbench thermal conductivity benchmark.
In this Section, we aim to provide a more comprehensive evaluation of Orb-v3 as well as other models from literature in terms of their ability to predict material properties -- beyond what is included in Matbench Discovery.
We believe this is important for scientists and engineers who wish to decide on which model they will use to fuel their computational research.

In addition to the Matbench suite of evaluations, we also consider the MDR phonon benchmark \cite{Togo2023, loew2024mdr, fu2025learning}, which presents a database of roughly ten thousand materials along with their vibrational and derived thermodynamic properties as computed at the PBE and PBEsol levels using \texttt{Phonopy}.
This benchmark is more comprehensive than the one included in Matbench since (1) its reference dataset is two orders of magnitude larger, and (2) it covers a wider range of physical observables depending on both the low- and high-frequency behavior of the material.
Second, we evaluate the models' ability to predict mechanical stability, based on a large subset of about ten thousand materials with precomputed PBE-level bulk and shear moduli from MP \cite{mp-pbe-elasticity-2025}.
These mechanical properties are complimentary to those obtained using (constant cell) phonon calculations, and the combination of these two benchmarks comprises a total of six physical properties.
Note that in the present evaluation, all six properties rely on finite difference estimates of higher-order PES derivatives and therefore require a MLIP to have a sufficiently smooth PES for successful evaluation.

\begin{table}[!t]
    \centering
\footnotesize
\begin{tabular}{lcccccc}
\toprule
 Property MAE & $\omega_{\max}$ & $S$ & $F$ & $C_V$  & $K_\text{bulk}$ & $K_\text{shear}$ \\
 Unit& [K] & [J/mol$\cdot$K] & [kJ/mol] & [J/mol$\cdot$K] & [GPa] & [GPa] \\
\midrule
MACE-MPA-0 [MPtraj+Alex] & 31 & 20 & 8 & 6 & 14 & 10 \\
eSEN-30M [MPtraj] & 21 & 13 & 5 & 4 & N/A & N/A \\
MACE-OMAT-0 & 17 & 10 & 3 & 3 & 13 & 9 \\
SevenNet-MF-ompa & 13 & 8 & 3 & 2 & 12 & 15 \\
\midrule
Orb-v3-conservative-inf-omat & 7 & 6 & 2 & 1 & 8 & 9 \\
Orb-v3-conservative-20-omat & 10 & 9 & 3 & 2 & 9 & 9 \\
Orb-v3-direct-inf-omat & 10 &  8 &  2 & 1 & 12 & 14 \\
Orb-v3-direct-20-omat & 11 &  10 & 3 & 2 & 12 & 16 \\
\bottomrule
\end{tabular}
\caption{Summary of the performance of current models across various physical property prediction benchmarks.
The first four columns cover both low- and high-frequency vibrational properties from the MDR phonon benchmark \cite{loew2024universal, Togo2023}; the highest phonon frequency $\omega_{\text{max}}$, the vibrational entropy $S$, free energy $F$, and heat capacity $c_\text{V}$.
The last two columns cover mechanical properties, and were obtained using \texttt{MatCalc} and the associated benchmark dataset of elastic constants \cite{matcalc}.
A full overview of all computational details is given in Appendix \ref{sec:computational_details}.
}
\label{mdr_phonon_table}
\end{table}

Table \ref{mdr_phonon_table} presents the performance of a variety of models across these properties, and it allows us to make two major observations.
First, \texttt{orb-v3-conservative-inf-omat} achieves the highest accuracy for almost all of the metrics in the table, while being faster than any of the best-performing models currently available in literature.
This is a clear demonstration that architectural constraints can be relaxed in the interest of performance, \emph{provided} that there is a sufficient amount of high quality QM data available to train on.
At present, this condition is evidently satisfied by the OMat24 dataset, which contains $\sim 55$ million AIMD-sampled structures.
The second observation is that even a non-conservative model with a sparse graph featurization such as \texttt{orb-v3-direct-20-omat} is comparable in accuracy to the current state of the art in literature.
This is remarkable, considering that it is about ~30 times faster than SevenNet, the current best performing model in literature (see Figure \ref{fig:profiles} for speed benchmarks).

\section*{Equigrad - Learned Rotational Invariance}
\label{sec:equigrad}

To incentivize learned invariance during training, we introduce \emph{equigrad}, a simple, differentiable metric which quantifies the degree of rotational invariance and which can be used as a regularization method during training.
Conceptually, we compute a gradient of the predicted energy $E$ with respect to an identity rotation matrix $\bm{R}$ that is inserted into the computational graph at the input.
An elegant way to achieve this is by first expressing an identity rotation $
\bm{R}$ as the matrix exponential of a skew-symmetric null matrix, and then computing the gradient of $E$ with respect to that null matrix:
\begin{align}
    \bm{R} = e^{\bm{G} - \bm{G}^T} \qquad \text{and} \qquad
    \Delta_\text{rot} = \frac{\displaystyle \partial E\left(\bm{r}^T \bm{R}, \bm{h} \bm{R}\right)}{\partial \bm{G}} \Bigg|_{\bm{G} = \bm{0}}
    \label{eq:equigrad}
\end{align}
where $\bm{r}$ are atomic positions and $\bm{h}$ is the cell matrix.

Invariant models have by definition $\Delta_\text{rot} = \bm{0}$ because the predicted energy does not depend on the global orientation of the input coordinates and cell vectors. For non-invariant models trained with data augmentation, $||\Delta_{\text{rot}}||$ is naturally small but nonzero, and quantifies the hypothetical change in energy if a rotation were to be applied at the input.

For conservative models, evaluation of Equation \ref{eq:equigrad} is essentially trivial since computing the interatomic forces and virial stress already require a backward pass through the network.
As such, we can apply L2-regularization to $\Delta_\text{rot}$ during training to incentivize rotational invariance of $E$ at no additional cost.

\begin{figure}[!t]
  \centering
  \begin{minipage}[c]{0.50\textwidth}
    \centering
    \raisebox{-0.5\height}{ 
      \includegraphics[width=0.95\linewidth]{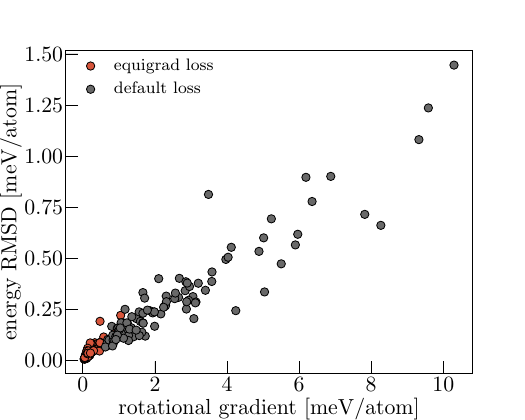}
    }
  \end{minipage}
  \hfill
  \begin{minipage}[c]{0.48\textwidth}
    \centering
    \raisebox{-0.5\height}{ 
      \footnotesize
      \begin{tabular}{lcc}
        \toprule
         & \multicolumn{2}{c}{$K_\text{SRME}$} \\
          & \makecell{\texttt{is\_plusminus} \\ \texttt{true}} & \makecell{\texttt{is\_plusminus} \\ \texttt{auto}} \\
        \midrule
            default loss & 0.222 & 0.868 \\
        equigrad loss & 0.232 & 0.365 \\
        \midrule
        MACE-MPA & 0.412 & 0.412 \\
        \bottomrule
      \end{tabular}}
  \end{minipage}
  
  \caption{\textbf{(left)} Scatter plot comparing the measured invariance (the standard deviation of the energy prediction over a randomized set of rotations) to the norm of the rotational gradient $||\Delta_\text{rot}||$, for all 103 structures in the thermal conductivity benchmark. Gray dots are obtained using Orb-v3 trained on OMat24 with the default loss function; red dots are obtained using Orb-v3 trained with equigrad regularization.
  \textbf{(right)} Thermal conductivity benchmark performance for two different methods in \texttt{Phonopy}; \texttt{auto} exploits the crystal symmetry to reduce the number of displacements to consider. For non-invariant models, this reduction is invalid, but models trained with equigrad regularization partially alleviate this difference due to increased invariance under rotation. }
  \label{fig:equigrad}
\end{figure}

Figure \ref{fig:equigrad} demonstrates the efficacy of equigrad in improving rotational invariance; the scatter plot on the left shows that the rotational invariance of Orb-v3 improves by $\sim$5x when training includes equigrad regularization. The table on the right demonstrates improved robustness of equigrad-trained models for crystal-symmetry-based workflows which make assumptions about equivariance, such as thermal conductivity calculations with \texttt{Phonopy}.

\section*{Uncertainty Estimates}
\label{sec: uncertainty}
Inspired by the widespread use of the per-residue 1DDT-C$\alpha$ (pLDDT) scores predicted by Alphafold \cite{alphafold2} as a confidence measure for structure prediction quality, we introduce a similar intrinsic binned confidence prediction for atomic force errors. All Orb-v3 models include a confidence head which predicts this binned atomic force error based on the final per-atom node representations.

\begin{algorithm}[htbp]
\caption{Per Atom Intrinsic Force Confidence}
\label{alg:confidence}

\textbf{perAtomForceConfidence}{$\{s_i\}, v_{\text{bins}} = [1, 3, 5, \dots, 50]^\top, \{r_i^{\text{Force MAE}}\}, c = 128$}
\begin{algorithmic}
    \State $a_i = MLP_{conf}(s_i)$
    \Comment $a_i, \text{ and intermediate activations} \in \mathbb{R}^{c}$
    \State $p_i^{ifc} = \text{softmax}(a_i))$
    \Comment $p_i^{\text{ifc}} \in \mathbb{R}^{|v_{\text{bins}}|}$
    \State $p_i^{\text{true ifc}} = \text{onehot}(r_i^{\text{true ifc}}, v_{\text{bins}})$
    \State $\mathcal{L}_{\text{conf}} = \text{mean}_i (p_i^{\text{true ifc}^\top} \log p_i^{\text{ifc}})$
    \State $r_i^{\text{ifc}} = \text{argmax}(p_i^{\text{ifc}})$
    \Comment $r_i^{\text{ifc}} \in v_{\text{bins}}$
    \State \textbf{return} $r_i^{\text{ifc}}, \mathcal{L}_{\text{conf}}$ \\
\end{algorithmic}
\end{algorithm}

\begin{figure}[!ht]
    \vspace{-1em}
    \centering
    \includegraphics[width=\linewidth]{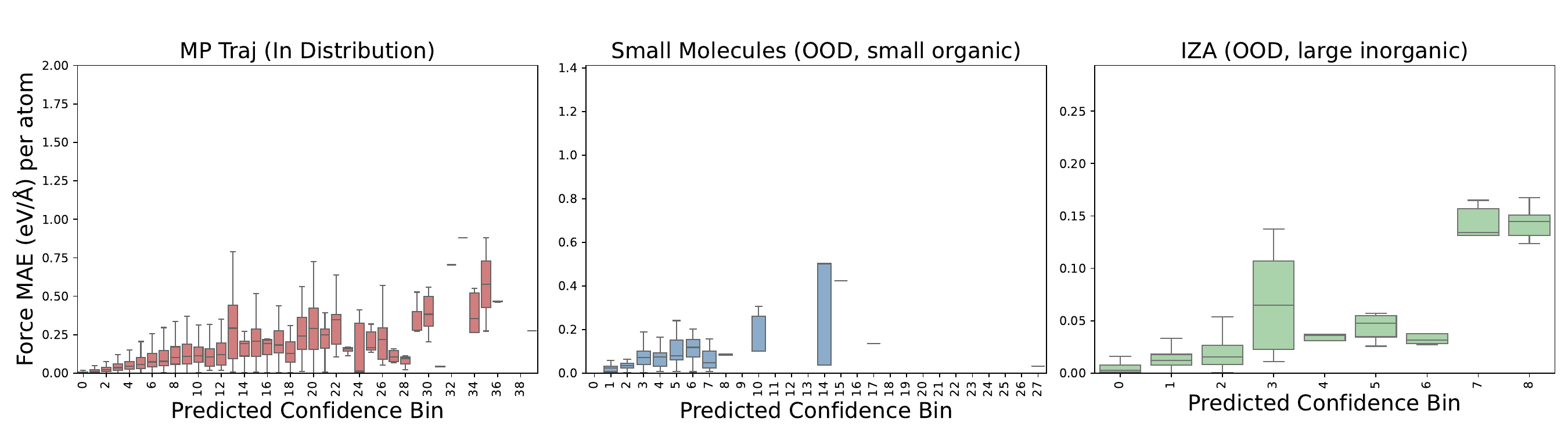}
    \caption{\small{Binned confidence predictions from Orb-v3's confidence head on on a random sample of systems from 3 datasets. MP Traj systems are sampled from the validation set; Small Molecules are systems randomly sampled from optimization trajectories of 162 commmon organic molecules from \cite{g2_ase_mols} (the g2 subset, made available in ASE), and IZA are 233 relaxed zeolite structures, all optimized with VASP at the PBE level of theory. Even for out of distribution datasets, confidence bin predictions correlate well with Force MAE at the atom level.}}
    \label{fig:confidence}
\end{figure}

To train the confidence head, we use the force errors produced by the model in an online fashion during model training. As such, the error distribution is dynamic, with error magnitudes decreasing as training progresses. In order to stabilize training on this shifting distribution, we train the confidence head using force predictions with a maximum error of 0.3 \r{A}, so as to provide a more calibrated confidence measure at distances which are more representative of a converged model's force predictions. Additionally, we use detached node representations from our model, meaning only confidence head parameters are affected by gradients from the confidence head loss.  Figure \ref{fig:confidence} shows that the intrinsic predicted confidence bin correlates well with force MAE, indicating that it may be useful for practitioners involved in active learning, data selection and other computational filtering workflows.

\section*{Conclusion}

We have presented the Orb-v3 family of interatomic potentials, which redefine the performance-speed-memory Pareto frontier for universal MLIPs. Our most significant achievement is the construction of extremely lightweight potentials that can model a variety of physical properties with an accuracy that matches or exceeds expensive, physically constrained models such as those in the MACE or SevenNet family \cite{batatia2024foundationmodelatomisticmaterials, park2024sevennet}.
In particular, our \texttt{orb-v3-direct-*-omat} models demonstrate how direct-force prediction reconciles accuracy and speed on established phonon prediction benchmarks while emphatically disproving the paradigm that conservatism and equivariance are strict prerequisites for universal MLIPs.

Across our publicly released models, we have introduced several features we hope will be useful to practitioners, such as substantial improvements in speed compared to Orb-v2, increased equivariance, and an intrinsic confidence measure. This confidence measure is inspired by the pLDDT scores predicted by Alphafold \cite{alphafold2} and we hope it has similar utility in enabling scientists to gain a visual insight into what the model does and does not "understand" on a per-atom basis. We are also excited by the potential of confidence measures to unlock new types of self-distillation and active learning \citep{tan2023single}.

A promising avenue for future work is to find a way to obtain the memory and speed benefits of neighbor limits without sacrificing any performance. The key question in our view is: how can we process fewer edges without losing too much information or inducing discontinuities in the PES? Taken to its extreme, this question suggests that \emph{edgeless} architectures may represent the future of ultra-efficient MLIPs, provided that they can be appropriately engineered to match the performance of edge-based GNNs.

\subsection*{The New Frontier: Meso-scale All-atom Simulations}
\label{section: mesocale_simulation}
Orb-v3's most obvious application is replacing DFT in conventional workflows with a more efficient method with comparable accuracy and lower memory requirements. However, this merely enhances rather than transforms our simulation capabilities.

Far more exciting is the possibility of applying Orb-v3 to study systems that have previously been  impossible to simulate accurately due to the large number of atoms involved and the lack of existing accurately parameterized empirical forcefields \cite{duignanPotentialNeuralNetwork2024}. Orb-v3 opens a new frontier where quantum mechanical accuracy can be maintained while exploring emergent phenomena arising from the collective behavior of thousands of atoms, such as crystal nucleation and growth \cite{zhangScalableAccurateSimulation2025}, self-assembly of complex nanostructures such as metal organic frameworks \cite{edwards2025exploring}, or phase diagrams of complex alloys \cite{zhuAcceleratingCALPHADbasedPhase2024}.

\begin{figure}[!ht]
    \centering
    \includegraphics[width=0.75\linewidth]{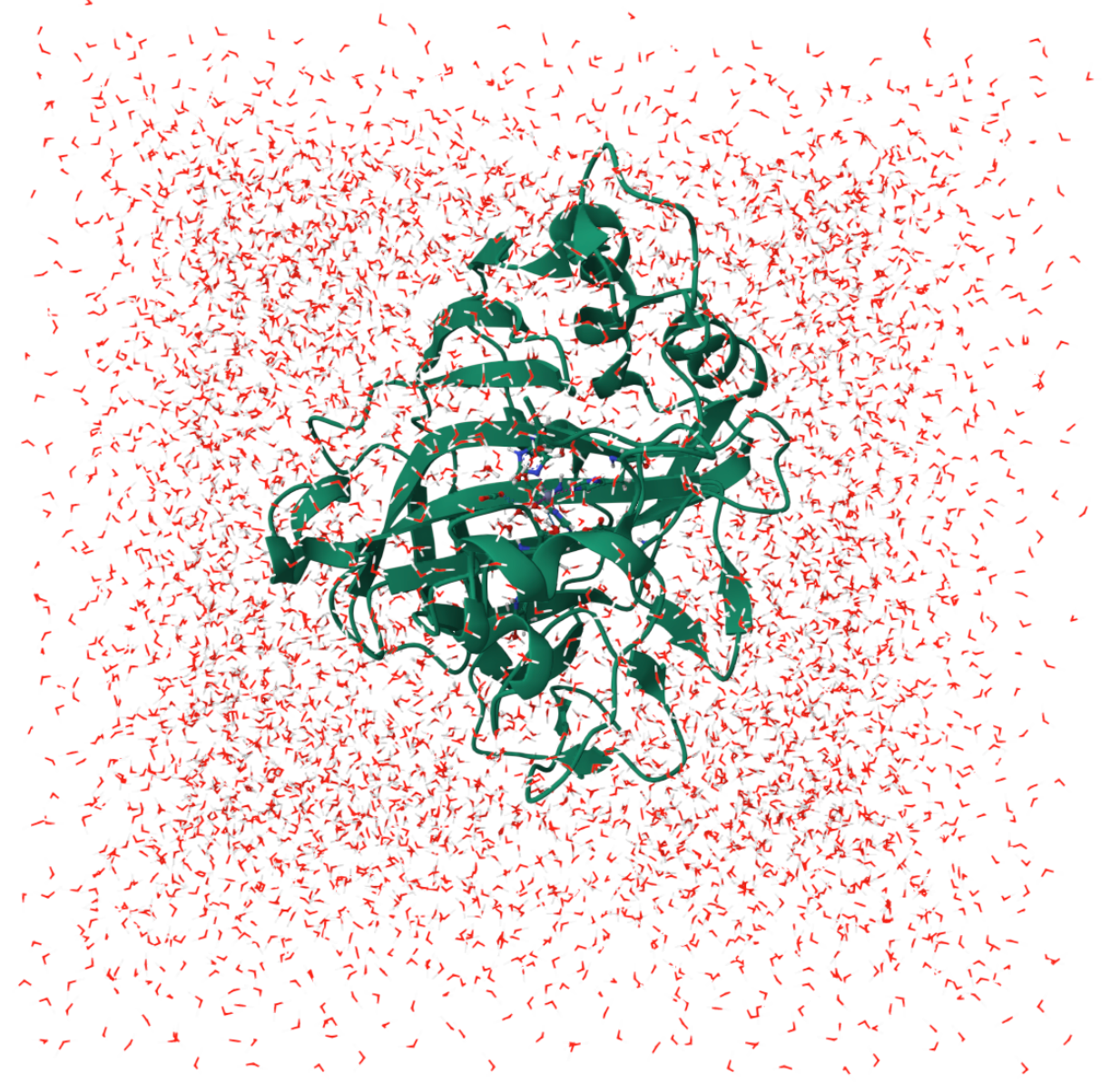}
    \vspace{-0.7em}
    \caption{Stable simulation of the Carbonic Anhydrase enzyme II system using \texttt{orb-v3-direct-inf-omat} for over 700 ps. The enzyme is depicted as its amino acid representation for visual clarity, but simulations use the full all-atom representation.}
    \label{fig:CAStructure}
\end{figure}

For example, in concurrent work \citep{duignan2025carbonic}, we have demonstrated the potential to study such mesoscale systems by simulating the carbonic anhydrase II enzyme. Using \texttt{orb-v3-direct-inf-omat} we simulate this enzyme under fully solvated conditions with no physical constraints using Langevin dynamics at 300 K. (See Figure~\ref{fig:CAStructure}). Despite being extremely out-of-distribution, and containing over 20,000 atoms, we do not observe unphysical behavior and the structure remains close to the original PDB structure throughout. 

While additional validation work remains to be done, the fact that Orb-v3 can provide long, stable simulations of a system so far outside the training data distribution is a strong indicator of the generality and potential of this new tool.  

\newpage
\printbibliography

\begin{appendices}
  \titleformat{\section}
    {\normalfont\Large\bfseries}
    {Appendix \thesection:}
    {1em}
    {}
\section{Code Availability}
\label{sec: code}

Model weights and code are available under an Apache 2.0 License on Github at \\ \url{https://github.com/orbital-materials/orb-models}.

\section{Lessons from Orb-v2}

\textbf{Successes.}
Orb-v2 \cite{neumann2024orb} was the first universal MLIP to demonstrate that a non-equivariant, non-conservative architecture can perform stable Molecular Dynamics (MD) on a range of out-of-distribution systems, whilst often obtaining qualitatively correct Radial Distribution Functions (RDFs) relative to the PBE \cite{Perdew1996GeneralizedGA} functional it was trained on. This achievement, combined with its superior speed compared to other universal MLIPs, and strong comprehensive benchmarking performance \cite{wines2024chips}, was a strong motivation for its continued development.

\textbf{Limitations.}
Several works \cite{pota2024thermal, loew2024universal} find Orb-v2 yields inaccurate finite-difference estimates of second and third order derivatives of the PES when using small atomic displacements, resulting in poor thermal conductivity estimates. \citet{zhao2025harnessing} observe that Orb-v2 underperforms many other potentials in identifying transition state pathways; again, this is a workflow involving higher-order information from the PES. The \texttt{MLIPX} benchmarking tool \cite{MLIPX} has revealed that Orb-v2's geometry optimizations of out-of-distribution slab-adsorbate systems can be unreliable with non-convergent energy graphs. Finally, a limitation has been highlighted by \citet{bigi2024dark}, who demonstrated that existing non-conservative models systematically fail to conserve energy in NVE MD simulations.

\textbf{Diagnosis.}The last two limitations primarily stem from non-conservatism. The other limitations are more subtle, but we have broadly arrived at the same conclusion as \citet{fu2025learning}, namely that \emph{enforcing smoothness} can be critical for downstream tasks involving higher-order derivatives of the PES. Unlike \citet{fu2025learning}---whose starting point was an Equiformer architecture \cite{liao2023equiformerv2}---our starting point of Orb-v2 is already relatively smooth due its use of a small number of radial basis functions and smooth envelope cutoffs in its attention layers. Nevertheless, we find room for improvement on this front, as captured in our modelling updates below.

\section{Orb-v3 modelling updates}
\label{sec: orbv3 modelling updates}
Motivated in part by the above limitations, as well as the desire for increased speed, Orb-v3 deviates in a significant number of ways from its predecessor:

\textbf{Model Compilation.} A simple but important update was to compile the model in PyTorch \cite{paszke2019pytorch}. Models are compiled by default whilst still allowing for dynamic graph sizes due to Pytorch's advanced compilation engine, which can take into account dynamic shapes. Importantly, Orb-v3 requires \texttt{torch==2.6.0} because there is a bug involving compilation of computation graphs containing \texttt{RMSNorm} in previous versions of torch.

\textbf{Width over depth}. We increase the width of every MLP in the GNS backbone from 512 to 1024. This allows us to train a 5 layer model with approximately the same parameter count ($\sim 25M)$ as Orb-v2, but is $2-3 \times$ faster.

\textbf{Direct and conservative models}. In addition to direct models, we also release conservative models that compute forces and stress via backpropagation of the energy with respect to positions and a symmetric displacement tensor, respectively \cite{langer2023stress, batatia2023macehigherorderequivariant}.

\textbf{Larger, more diverse dataset.} Our main models are trained on OMat24 (AIMD only), rather than the Mptraj and Alexandria datasets used by Orb-v2.
    
\textbf{Smoother edge embeddings.} 
The edge embeddings in Orb-v2 were a concatenation of each edge vector (normalized to unit length) and 20 Gaussian radial basis functions (RBFs) applied to the edge length. In Orb-v3, we instead compute an outer-product between Bessel radial basis functions and Spherical Harmonic angular embeddings. Specifically we use 8 Bessel bases and set $L_{max}=3$ for the spherical harmonics.

\textbf{Huber loss and pair repulsion.} 
We adopt two useful ideas from \citet{batatia2024foundationmodelatomisticmaterials}. Firstly, we switch from using mean absolute error losses for energies, forces and stress, to using Huber losses (delta = 0.01). We also include a non-learnable ZBL pair repulsion term in our models, enabling them to more accurately model strong repulsive forces for atoms close together.

\textbf{Controllable max neighbors.} We release models with unlimited numbers of neighbors, in addition to a maximum of 20 as used by Orb-v2. As demonstrated throughout the paper, limiting neighbors reduces costs; both the graph construction cost and the cost of the model forward pass. It does however induce subtle discontinuities in the PES, which induces a modest performance penalty for certain workflows.

\textbf{Confidence  Head}. Inspired by Alphafold's \cite{alphafold2} per-residue 1DDT-C$\alpha$ (pLDDT) scores, we add a confidence head to Orb-v3 which produces an intrinsic binned confidence measure. See main text for full explanation.

\subsection*{Workflow considerations} 

Several common computational chemistry workflows implicitly assume either strict conservatism or roto-equivariance. For instance, line-search-based optimization algorithms assume strict energy-force consistency and \texttt{Phonopy}’s \texttt{is\_plusminus=`auto'} displacement generator assumes strict roto-equivariance. It is important for users to be aware of these assumptions, and consider alternative approaches in order to obtain the best performance when using non-invariant, non-conservative models.

In the case of \texttt{Phonopy}'s displacement generator, its default behavior is to exploit rotational/translational symmetry of crystal space groups in its finite difference approximations, which is mathematically invalid when using a non-invariant potential.

Fortunately, these limitations can often be sidestepped via a more rigorous choice of settings (\texttt{is\_plusminus=True} in \texttt{Phonopy}) or alternative algorithms (non-line-search based optimizers such as FIRE). When no workaround is possible, as may be the case for strict energy conservation in NVE molecular dynamics, then we recommend using more architecturally constrained models, like \texttt{orb-v3-conservative-inf}.

\section{Efficient graph construction}
\label{sec: efficient graph construction}

\begin{figure}[!ht]
    \vspace{-1em}
    \centering
    \includegraphics[width=\linewidth]{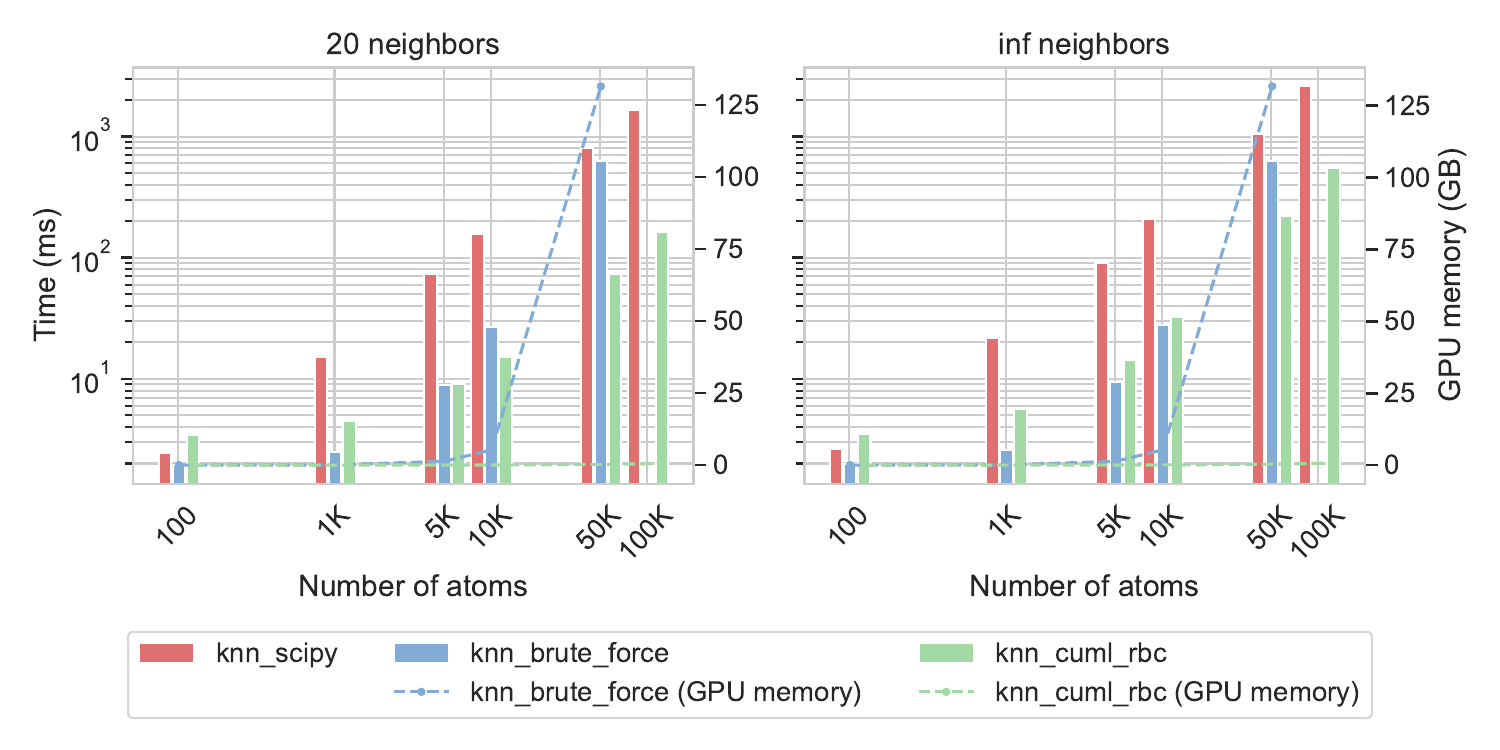}
    \caption{Timing (left axis) and GPU memory use (right axis) for a variety of KNN graph creation for varying periodic system sizes and number of neighbors. Of particular note is the \emph{cuml} library, which includes memory efficient graph construction methods for nearest neighbors computatation on GPU.}
    \label{fig:featurization-benchmark}
\end{figure}

Figure \ref{fig:featurization-benchmark} shows a variety of graph construction methods:

\begin{itemize}
    \item \texttt{scipy.spatial.KDTree} - A CPU only implementation of a kd-tree. \cite{kdtree_scipy}
    \item \texttt{Brute Force (torch.cdist, torch.topk)} - matrix multiplication based nearest neighbors, where all pairwise distances are computed, before the topk are selected. This is extremely memory intensive with a lot of wasted computation for large systems. However, as the problem is embarrassingly parallel, this can work effectively in practice.
    \item \texttt{cuml.neighbors.NearestNeighbors(algorithm="rbc")} - GPU accelerated ball tree implementation of nearest neighbors.
\end{itemize}

Figure \ref{fig:featurization-benchmark} demonstrates that for consistently good performance across a variety of system sizes, the graph construction method must be \textit{adaptive}. For small system sizes, the overhead of GPU based graph construction is too high; for slightly larger system sizes, brute force matrix multiplication based GPU routines offer the best performance, and at very large system sizes, memory considerations require the use of a combination of GPU acceleration and algorithmic efficiency. In certain scenarios, for mesoscale simulations, a practitioner may come full circle, choosing nearest neighbor implementations which are CPU compatible (at the cost of performance), in order to relieve pressure on accelerator memory - this can again change the equation for which method is optimal for a given simulation.

Orb-v2's graph featurization used  a fixed ($3\times3\times3$) supercell expanded from a central unit cell. The correctness of this approach depends on the max neighbors, radius cutoff and  the size of the minimum unit cell dimension. Instead, we now construct the supercell dynamically, computing the minimum number of unit cell tilings in a given cell direction to ensure correct graph construction.

\section{Energy conservation}
\label{sec:energy_conservation}
\begin{figure}[t]
    \centering
    \includegraphics[width=.6\linewidth]{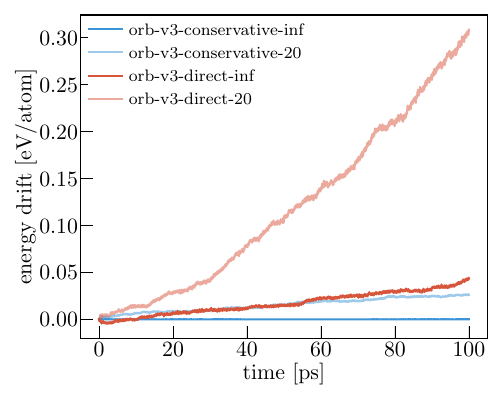}
    \caption{Total energy during NVE dynamic simulations as a function of time, for the various Orb-v3 models.
    Only \texttt{orb-v3-conservative-inf} is truly energy-conserving, so this model is to be recommended whenever calculating physical properties based on constant energy dynamics.}
    \label{fig:nve}
\end{figure}

While most experimental observables are predicted from simulations that are performed at constant temperature and/or pressure, there are some workflows which rely on constant energy dynamics.
In those scenarios, it is important to evolve the dynamics of the system using continuous and conservative forces.
Within Orb-v3, the only model that satisfies these constraints rigorously is \texttt{orb-v3-conservative-inf}, and Figure \ref{fig:nve} demonstrates this for an arbitrary system in the MPtraj dataset.
While \texttt{orb-v3-conservative-20} still computes the forces as gradient of the energy, the neighbor limit per atomic environment implies that small discontinuities are going to be present, and these give rise to non-energy-conserving behavior.
The \texttt{orb-v3-direct-inf} model does exhibit rigorously continuous forces but as they are not computed as the gradient of a scalar, they are non-conservative.
Finally, \texttt{orb-v3-direct-20} is both non-conservative and exhibits small discontinuities in the forces, and this naturally gives rise to the largest energy drift.

\section{Thermal conductivity calculations}
\begin{figure}[t]
\centering
\includegraphics[width=0.7\linewidth]{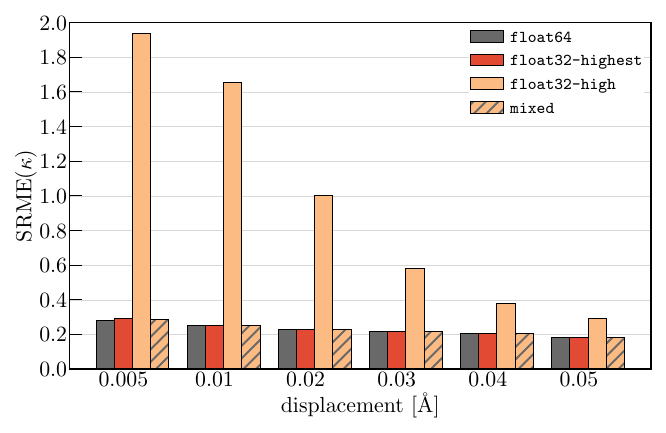}
\caption{Variation of the evaluated $\kappa_{\text{SRME}}$ with displacement step size as used by \texttt{Phonopy} to estimate the second- and third-order derivatives -- for different \texttt{PyTorch} precision levels.
The hatched bar refers to an experiment using low precision for the geometry optimization (\texttt{float32-high}) but a high precision for the subsequent finite difference evaluations (\texttt{float64}).
}
\label{fig:ksrme}
\end{figure}

We observe that the prediction error for thermal conductivities (as measured by the SRME) is somewhat dependent on the step size used by \texttt{Phonopy} in its finite difference approximation to the higher-order derivatives of the PES; this has been reported by other authors as well \cite{fu2025learning}.
In addition, the evaluation is observed to depend on the floating point precision used to evaluate the forces -- see Figure \ref{fig:ksrme}.
To identify exactly which part of the calculation is causing this, we ran a mixed precision experiment in which the geometry relaxation is performed in low precision while the subsequent force evaluations are performed in high precision.
Figure \ref{fig:ksrme} shows that this approach achieves essentially the same accuracy as running the whole experiment in high precision, which indicates that the loss in accuracy at reduced precision is \emph{not} related to failures in the geometry optimizations but instead relates to a breakdown of the finite difference approximations whenever forces are evaluated in low precision.

\section{MDR benchmark and mechanical properties}
\label{sec:computational_details}
This Section gives an overview of the computational details that are involved in the evaluation of models on the phonon MDR benchmark and on the mechanical property benchmark (Table \ref{mdr_phonon_table}).

For the phonon MDR, we use \texttt{Phonopy} to generate displacements and compute the (second-order) force constants.
Before applying displacements, atomic positions and unit cell components are first optimized using a combination of the FIRE optimizer and a \texttt{FrechetCellFilter} from the Atomic Simulation Environment (ASE) \cite{larsen2017ase}.
We use a displacement magnitude of 0.01 \r{A} and \texttt{is\_plusminus=True} to generate displacements, and a default $q$-mesh of [20,20,20].
Free energy, entropy, and heat capacity were evaluated at 300 K based on the obtained force constants.

For the bulk and shear moduli, we sub-sampled 1,000 materials from the full benchmark datasets to limit the total time required for its evaluation.
Before applying the strain displacements, atomic positions and unit cell components were optimized using a combination of the FIRE optimizer and a \texttt{FrechetCellFilter} from the Atomic Simulation Environment (ASE) \cite{larsen2017ase}.
We use strain magnitudes of [-0.1, -0.05, 0.05, 0.1] for the normal (diagonal) components, and [-0.02, -0.01, 0.01, 0.02] for the off-diagonal components as we found this to yield the best agreement with the PBE reference values across all models (though it is possible that there is some level of error cancellation involved here).
After applying strain to the optimized unit cell, atomic positions were optimized at fixed unit cell, as per the original MP protocol.

\section{Distillation for direct models}
\label{sec: distillation direct mpa}

As stated in the main text, we find that distillation-based training with conservative teachers promotes more accurate force-derivatives for our direct \texttt{mpa} models. Such distillation is not required when training direct models on \texttt{omat}, suggesting that some unique quirk of the \texttt{mpa} force distribution causes degradation (and this quirk is absent in the conservative model predictions we distill from).

Identifying the exact nature of this "quirk", and understanding whether or not it exists in other datasets is an important topic for future research. If the degradation of direct forces is a common occurrence across a range of downstream finetuning datasets, then improved forms of distillation may become essential. The distillation method used in this work is rather basic and does not make use of new, hessian-based methods for MLIPs \cite{hessian_finetune_aditi, rodriguez2025doeshessiandataimprove}.

\section{Effect of filtering OMat24}
\label{sec: filtering omat}

\begin{figure}[!ht]
    \centering
    \includegraphics[width=\linewidth]{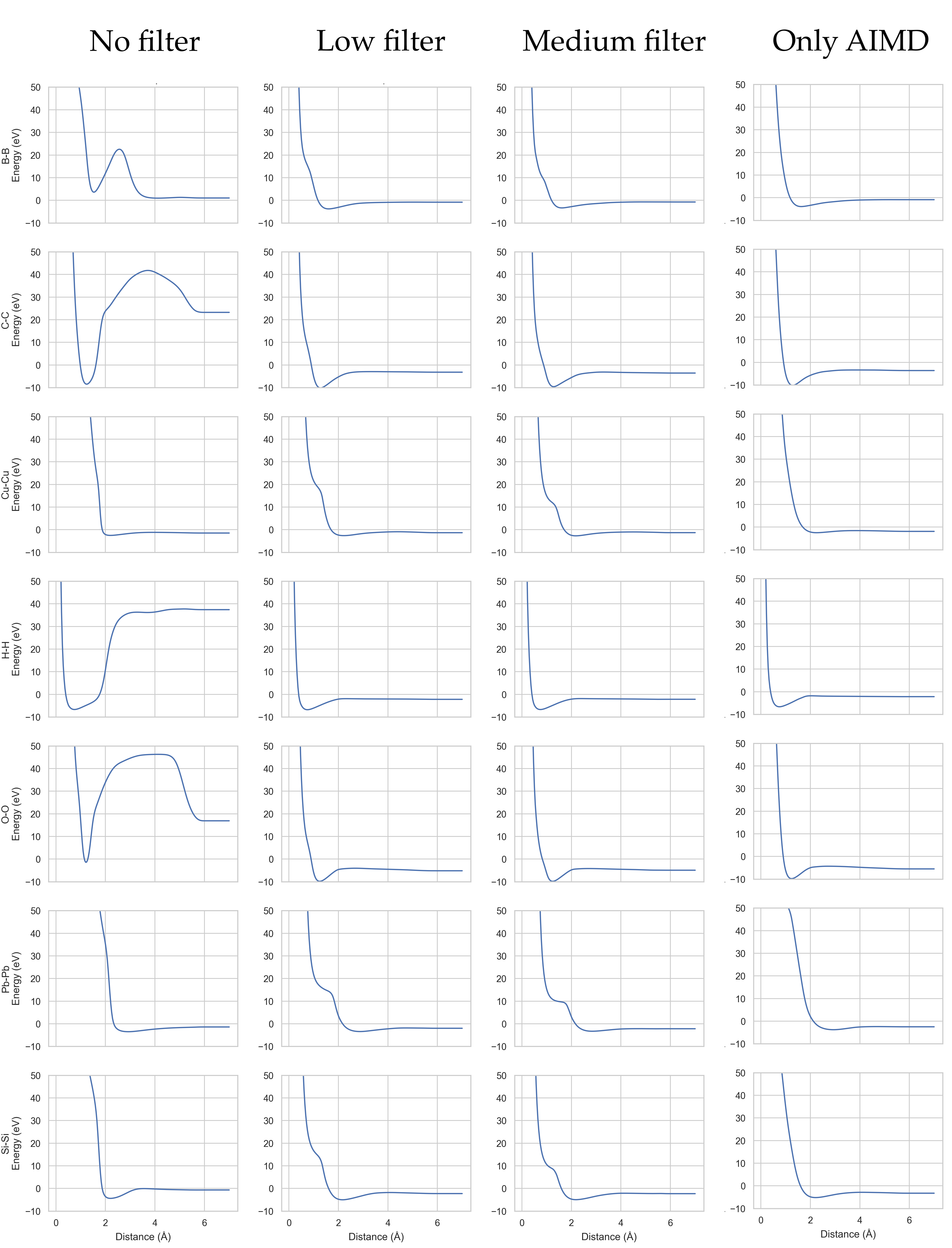}
    \caption{\small{Diatomic energy curves for conservative 5-layer Orb-v3 models trained on different versions of the OMAT24 dataset. The leftmost column uses the full OMAT24 dataset for training without any filtering. The "low filter" removes all datapoints with energies above 10 eV, maximum atomic force above 50 eV/\r{A} and maximum eigenvalue of the stress matrix above 1.0 eV/$\text{\r{A}}^3$; this removes a total of 0.4\% of the dataset. The "medium filter" applies more aggressive filtering with thresholds of 0.0 eV, 30 eV/\r{A} and 0.3 eV/$\text{\r{A}}^3$, thereby removing 2.8\% of the dataset. The final column only uses the AIMD subset of the OMAT dataset, discarding all "rattled" systems, which account for 45\% of the data.}}
    \label{fig:diatomics-omat}
\end{figure}

During development of the Orb-v3 potentials, we observed that all models (conservative or direct) suffered from undesirable out-of-distribution behavior when trained on the full OMat24 dataset and evaluated on homo-nuclear diatomics, as shown in the far left column of Figure \ref{fig:diatomics-omat}. Interestingly, models with such diatomics still had low $\kappa_{\text{SRME}}$ values for small bulk crystals, indicating that that this was not a general pathology across all systems, but emerged in the OOD setting of a two-atom system with one edge per atom.

Also shown in Figure \ref{fig:diatomics-omat} are different attempts to filter the OMat24 dataset. The central two columns show different amounts of filtering based on outlying energies, forces and stress. Such filtering was strongly beneficial, but still insufficient as large kinks remained in the energy surface. The only completely effective strategy that we tried was to remove all non-AIMD data, as depicted in the far right column.

Arguably, this is a dissatisfying outcome as we would like to avoid discarding valid DFT data. Whilst we broadly in favour of retaining as much of a model's training data as possible, it remains unclear if the large proportion of "rattled" systems in OMat24 (45\% of the data), and the amount by which they are rattled, is generally beneficial or not for the current generation of universal MLIPs, or whether the problems we have observed are unique to more unconstrained architectures.

\section{Compatibility between VASP pseudo-potentials}
\label{sec:compatability}
\begin{figure}[!ht]
    \centering
    \includegraphics[width=0.75\linewidth]{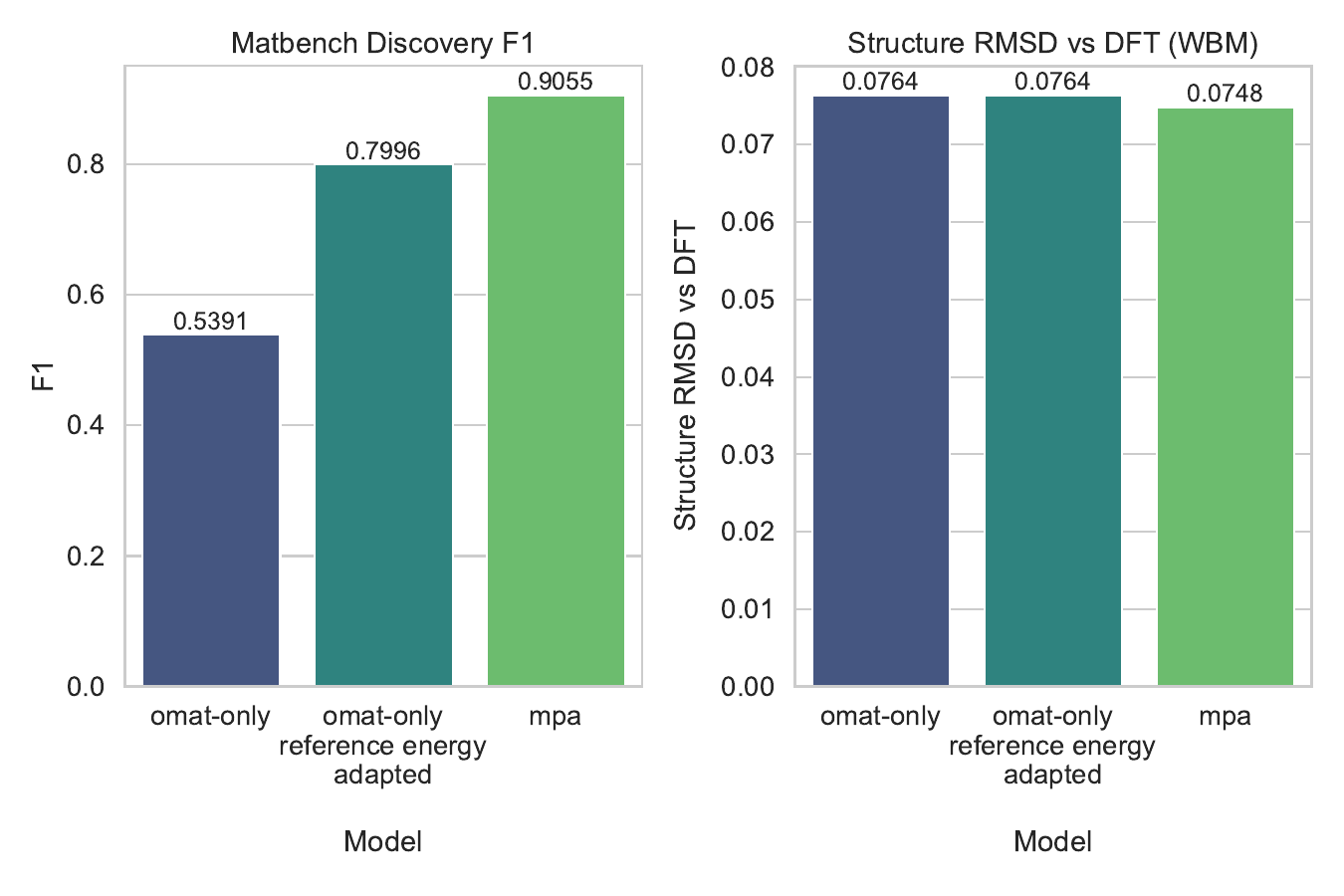}
    \caption{Matbench F1 and RMSD of optimizations on the WBM test set. There is a substantial increase in F1 (0.54 -> 0.80) for models trained on OMat24, but with re-initialized reference energies based on the coefficients of a least squares regressor fit to the MP-Traj. }
    \label{fig:pseudopotential_adaptation}
\end{figure}

Training methods which use OMat24 as either a pretraining step, or for joint training when evaluating on the Matbench datasets, have become more common due to its empirical impact on performance, despite the fact that the datasets are generated with incompatible pseudopotentials (PBE 52 and 54 respectively). In order to probe the differences in these pseudopotentials, we plot the difference between 3 model variants in Figure \ref{fig:pseudopotential_adaptation}. Firstly, models trained on OMAT only result in successful optimizations on the WBM dataset (the test set used for Matbench Discovery) when measured using RMSD. Secondly, we re-initialize the reference energies used in this OMAT base model to the coefficients of a least squares regressor fit to MP-Traj energies using atomic composition as features. This model sees a substantial boost F1 performance despite a marginal change in RMSD, suggesting that a constant factor shift in atomic energies can explain ~$70\%$ of the change in F1. In combination, these results suggest that the transfer between these two datasets can be explained by the fact that the gradient fields of the potential are very similar (they result in similar optimizations). Methods which finetune a small amount on MP-Traj are effective in large part because they are adjusting to a new energy distribution - despite $70\%$ of this variation being captured by a linear transformation with respect to chemical composition. 

This discrepancy highlights a difficulty in MLIP evaluation; combining new, incompatible datasets to achieve results on static benchmarks risks incentivizing methods for combining datasets which do not lead to more effective or performant models, such as very short post-training finetuning to adjust a model to a benchmark.

\section{Pareto Frontier Model Families}
\label{sec:model_families}

Model families in Figure \ref{fig:pareto} are composed of:

\begin{itemize}
    \item MACE
    \begin{itemize}
        \item \texttt{MACE-MP-0}
        \item \texttt{MACE-MPA-0}
    \end{itemize}
    \item SevenNet
    \begin{itemize}
        \item \texttt{7net-mf-ompa}
        \item \texttt{7net-l3i5}
        \item \texttt{7net-0}
    \end{itemize}
    \item Orb-v3 - All Orb-v3 variants described in the Models Section.
    \item MatterSim
    \begin{itemize}
        \item \texttt{Mattersim-v1.0-5M}
    \end{itemize}
\end{itemize}

\end{appendices}

\end{document}